\documentclass[%
 prl,
 superscriptaddress,
 amsmath,amssymb,
reprint,%
longbibliography,
]{revtex4-1}
\usepackage{hyperref}
\usepackage{amssymb}
\usepackage{aeguill}
\usepackage{graphicx}
\usepackage{dcolumn}
\usepackage{bm}

\usepackage{color}
\usepackage{url}
\usepackage{verbatim}
\usepackage{lineno}
\usepackage[utf8]{inputenc}
\usepackage{color}
\usepackage{lineno}
\usepackage{mathtools}

\usepackage{booktabs, caption}
\usepackage{siunitx}
\draft 

\begin{document}



\title{Mastering disorder in a first-order transition by ion irradiation} 



\author{S.~Cervera}
\affiliation {Institut des NanoSciences de Paris, CNRS, Sorbonne Université, F-75005 Paris, France}
\author{M.~LoBue}
\affiliation {SATIE, ENS Paris Saclay, CNRS, Université Paris-Saclay, F-91190 Gif-sur-Yvette, France}
\author{E.~Fontana}
\homepage{Present address: Institut Néel, CNRS, F-38042, Grenoble, France}
\affiliation {Institut des NanoSciences de Paris, CNRS, Sorbonne Université, F-75005 Paris, France}
\affiliation {Department of Applied Science and Technology, Politecnico di Torino, I-10129 Torino, Italy}
\author{M.~Eddrief}
\affiliation {Institut des NanoSciences de Paris, CNRS, Sorbonne Université, F-75005 Paris, France}
\author{V.H.~Etgens}
\thanks{Present address: Centre de recherche et de restauration des musées de France, F-75001 Paris}
\affiliation {Institut des NanoSciences de Paris, CNRS, Sorbonne Université, F-75005 Paris, France}
\author{E.~Lamour}
\affiliation {Institut des NanoSciences de Paris, CNRS, Sorbonne Université, F-75005 Paris, France}
\author{S.~Macé}
\affiliation {Institut des NanoSciences de Paris, CNRS, Sorbonne Université, F-75005 Paris, France}
\author{M.~Marangolo}
\affiliation {Institut des NanoSciences de Paris, CNRS, Sorbonne Université, F-75005 Paris, France}
\author{E.~Plouet}
\affiliation {Institut des NanoSciences de Paris, CNRS, Sorbonne Université, F-75005 Paris, France}
\author{C.~Prigent}
\affiliation {Institut des NanoSciences de Paris, CNRS, Sorbonne Université, F-75005 Paris, France}
\author{S.~Steydli}
\affiliation {Institut des NanoSciences de Paris, CNRS, Sorbonne Université, F-75005 Paris, France}
\author{D.~Vernhet}
\affiliation {Institut des NanoSciences de Paris, CNRS, Sorbonne Université, F-75005 Paris, France}
\author{M. Trassinelli}
\email[]{martino.trassinelli@insp.jussieu.fr}
\affiliation {Institut des NanoSciences de Paris, CNRS, Sorbonne Université, F-75005 Paris, France}

\date{\today}

\begin{abstract}
The effect of ion-irradiation on MnAs single crystalline thin films is studied. The role of elastic collisions between ions and atoms of the material is singled out as the main process responsible for modifying the properties of the material. Thermal hysteresis suppression, and the loss of sharpness of the magneto-structural phase transition are studied as a function of different irradiation conditions. While the latter is shown to be associated with the ion-induced disorder at the scale of the transition correlation length, the former is related to the coupling between disorder and the large-scale elastic field associated with the phase coexistence pattern.

\end{abstract}


\maketitle 


\section{Introduction}

The role of quenched disorder on phase changes has been addressed theoretically for several decades \cite{Harris1974-1, Imry1979-1, Sethna1993-1, Berger2000-1, Kirkpatrick2016-1}. 
In the case of first-order transitions, unraveling the leading physical mechanisms responsible for the coupling between random impurities and order parameters is even  harder than in second order ones.
Indeed, the coexistence of different characteristic lengths and the presence of non-equilibrium features like hysteresis and kinetics make the task even rather challenging.
Nonetheless, as most of the known large caloric effects take place on the verge of a first-order ferroic transition \cite{Moya2014}, mastering hysteresis and preserving other features (i.e. entropy and temperature changes) has been the focus of many investigations for the last two decades. 
At stake, beside the fundamental interest, such investigations can have a direct impact on the use of caloric effects for refrigeration and energy conversion applications \cite{Gutfleisch2016, Lyubina2017, Gottschall2019}. 
Hysteresis reduction in magnetocaloric materials has been successfully achieved by acting on the phase transition globally, getting closer to a critical point through chemical change 
\cite{Bruck2012-1, Franco2018}, using a secondary external field 
\cite{Liu2012,Mosca2008,Moya2013}, or 
using local non-homogeneity 
fostering phase nucleation. Indeed, controlling first-order phase changes in caloric materials using defects and disorder has been studied at different scales, from the atomic/chemical \cite{Sokolovskiy2012-1, Fujita2014-1}, to the structural one \cite{Lyubina2011-1,Niemann2016}. Therefore, the possibility of tuning the amount of disorder in a model system represents a unique opportunity for experimental investigations on phase change in solid-state systems.
Moreover, it can pave the way for developing new methods for tailoring the functional properties of materials where phase changes take place.

In this paper, we study the modifications induced by the collision and implantation of ions in manganese arsenide thin monocrystals.
Bulk manganese arsenide (MnAs) shows a first-order magneto-structural transition from ferromagnetic to paramagnetic order accompanied by a lattice symmetry switch when getting above the transition temperature $T_t = 313$~K \cite{Menyuk1969-1}. From this perspective, MnAs can be considered as a prototypical first-order magnetocaloric material. 
Previous studies \cite{Trassinelli2014,Trassinelli2017} proved thermal hysteresis suppression in ion-irradiation MnAs films, keeping the caloric effect intensity unchanged. 
Here the mechanisms underlying the modifications of the transition after ion-irradiation are investigated on single crystalline $150$~nm thick MnAs films under different irradiation conditions (i.e. ion mass, ion kinetic energy, fluences, etc.).  

\section{Experimental methods}

\subsection{Growth conditions}

The MnAs films have been produced by molecular beam epitaxy on GaAs(001) $0.3$~mm thick substrates with  
the $\alpha$-MnAs$[0001]$ 
axis parallel to GaAs$[\bar110]$ \cite{Breitwieser2009-1}.  
Due to the epitaxial strain,  the phase coexistence between the ferromagnetic $\alpha$-phase with hexagonal structure 
(NiAs-type) 
and the paramagnetic $\beta$-phase with orthorhombic structure, (MnP-type), 
takes place over an extended temperature range ($280$--$320$~K) \cite{Daweritz2006}, in contrast with the sharper transition observed at the temperature $T_t = 313$~K in bulk single crystals \cite{Wilson1963-1}. This widening of the phase-coexistence region is characterized  by a linear evolution of the phase fraction, across the transition, as a function  of $T$. As shown in  \cite{Kaganer2000-1, Kaganer2002-1, Plake2002-1, Plake2003-1, Daweritz2003-1}, this behavior is driven by a mesoscopic phase coexistence pattern often referred to as  stripes-domain pattern \cite{Breitwieser2009-1}, and by its long-range elastic field. 

\subsection{Irradiation conditions}

Ion irradiation has been performed at two facilities: the electron-cyclotron ion source SIMPA facility (Paris, France) \cite{Gumberidze2010}, and the line IRRSUD facility at the GANIL accelerator (Caen, France). 

At SIMPA, samples have been irradiated with different ions (helium, oxygen, neon, argon and krypton) with kinetic energies ranging from $22$ to $260$~keV. 
The ion fluence $\Phi$ has been varied between $1 \times 10^{12}$~ions/cm$^2$ and $6\times 10^{15}$~ions/cm$^2$, corresponding to an irradiation time spanning from few tens of seconds up to several hours. 
All irradiation have been performed at room temperature.
The energy of the different ion beams and the incident angle have been chosen in order to reach an average penetration depth corresponding to the half thickness of the film. 

\begin{table*}
\caption{\label{tab:data} Summary of the irradiation characteristics ordered by the density of the induced collisions. Fluence and flux values indicate the corresponding average value during the irradiation.}
\begin{ruledtabular}
\begin{tabular}{*{2}{S[table-format=1.2e-2]} cccc *{2}{S[table-format=1.2e-2]}}
{Coll. density} & {Ion density} & Ion & At. mass & {Kin. energy} & {Incident angle} & {Fluence}& {Flux} \\
{(cm$^{-3}$)} & {(cm$^{-3}$)} &  &  & {(keV)} & {($^\circ$)} & {(ions cm$^{-2}$)} & {(ions cm$^{-2}$ s$^{-1}$)} \\
\midrule
4.83E+19 & 0.00E+00 & Ar & 36 & 7800 & 0 & 4.00E+12 & 6.67E+09 \\
6.60E+19 & 1.97E+16 & Ar & 40 & 208 & 60 & 3.64E+11 & 2.43E+10 \\
1.85E+20 & 1.49E+18 & He & 4 & 22 & 60 & 3.27E+13 & 6.54E+12 \\
2.41E+20 & 0.00E+00 & Ar & 36 & 7800 & 0 & 2.00E+13 & 6.06E+09 \\
4.71E+20 & 9.17E+16 & Kr & 84 & 260 & 0 & 1.40E+12 & 9.33E+10 \\
7.18E+20 & 6.70E+17 & O & 16 & 84 & 60 & 1.34E+13 & 1.12E+12 \\
9.37E+20 & 6.54E+17 & Ne & 20 & 89 & 60 & 1.32E+13 & 1.96E+11 \\
1.07E+21 & 7.44E+17 & Ne & 20 & 90 & 60 & 1.50E+13 & 2.50E+11 \\
1.21E+21 & 0.00E+00 & Ar & 36 & 7800 & 0 & 1.00E+14 & 4.57E+09 \\
2.69E+21 & 5.24E+17 & Kr & 84 & 260 & 0 & 8.00E+12 & 8.00E+10 \\
3.47E+21 & 1.04E+18 & Ar & 40 & 208 & 60 & 1.91E+13 & 3.19E+11 \\
6.31E+21 & 1.89E+18 & Ar & 40 & 208 & 60 & 3.48E+13 & 2.32E+10 \\
7.66E+21 & 5.34E+18 & Ne & 20 & 89 & 60 & 1.08E+14 & 1.80E+11 \\
3.26E+22 & 2.63E+20 & He & 4 & 22 & 60 & 5.76E+15 & 6.40E+12 \\
4.82E+22 & 9.36E+18 & Kr & 84 & 260 & 0 & 1.43E+14 & 1.19E+11 \\
1.56E+23 & 4.65E+19 & Ar & 40 & 208 & 60 & 8.59E+14 & 5.72E+10 \\
\end{tabular}
\end{ruledtabular}

\end{table*}

At IRRSUD a beam of isotopic $^{36}\!$Ar with an energy of $35.28$~MeV has been used. In this case, no ions are implanted in the MnAs film, with most of them being stopped in the GaAs substrate. The fluence has been varied from $4 \times 10^{12}$ to $5\times 10^{15}$~ions/cm$^2$. 
The complete information about the irradiation conditions can be found in Table~\ref{tab:data}.

During the interaction with a solid the ion transfers its kinetic and potential energies to the target atoms by a series of processes that can be schematically separated into two categories: one due to the interaction with the target nuclei, and the other due to the interaction with the electrons. 
The electron interaction is inelastic, and entails small or no lattice distortion into the target. 
The ion energy losses, including those related to the potential energy of the ion, 
are transferred to the solid via excitation of phonons, with an increase in temperature along the trajectory. 
Nuclear collisions are the most probable process at low velocity and/or for heavy ions
\cite{Nastasi,Nordlund2018,Nordlund2019}. 
Such collisions lead to the displacement of sample atoms from their site and, if the kinetic energy is sufficiently high, to a cascade of secondary collisions, with the creation of clusters of vacancies and interstitial atoms in the sample lattice
\cite{Nastasi}.
For the same kinetic energy, heavy ions produce a higher number of collisions per ion than light ions.
This is due to the higher cross-section of the primary impact and the more favorable momentum transfer to target atoms, as well as the consequent secondary collisions in the cascade.

At the ion kinetic energies of SIMPA ($E_\text{kin}$ less than 300~keV), the main process responsible for induced modifications is the nucleus--nucleus interaction between incoming ions and sample atoms.
At the ion kinetic energies of IRRSUD, the probability of nuclear processes decreases while inelastic processes start to occur.
Due to the small ion potential energy ($E_p \lesssim 5$~keV), compared to the kinetic energy, and to the metallic nature of the MnAs samples, the local heating from the electron interaction is not expected to induce noticeable modifications in the material. 
Indeed, elastic nuclear collisions remain the leading process that modifies the properties of the irradiated sample.

\section{Results and discussions}

\subsection{General considerations}

\begin{figure}[h!]
\includegraphics[width=\columnwidth]{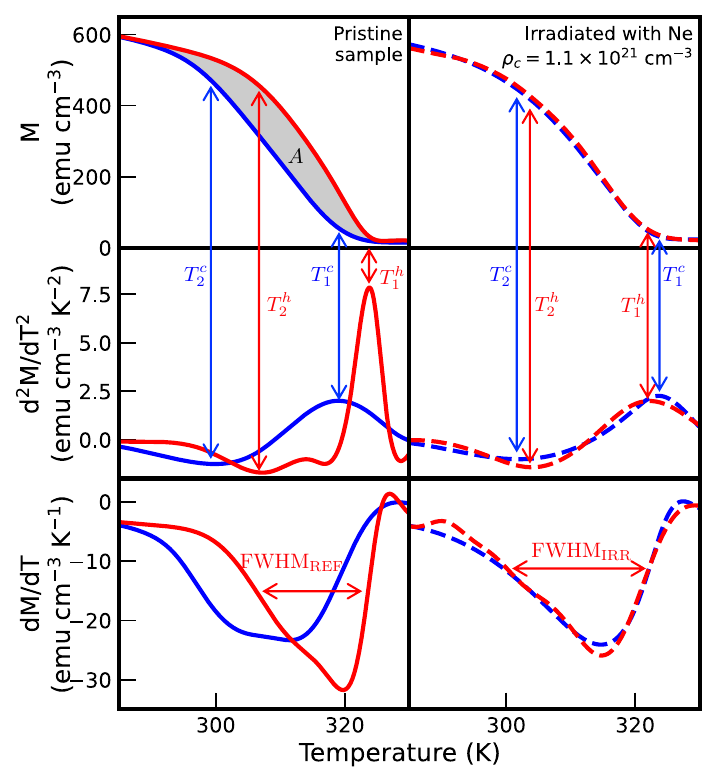}
\caption{Magnetization curves $M(T)$ (top) and corresponding first (bottom) and second derivative (middle) measured on reference (left) and irradiated (right) samples. Examples of $\mathrm{FWHM}_\mathrm{REF}$, and of $\mathrm{FWHM}_\mathrm{IRR}$, used in Eq.(\ref{eq:delta_T_rounding}), are depicted in the bottom frame left and right, respectively.
The temperatures $T_1^h$, and $T_1^c$ corresponding to the maximum of the second derivative $d^2M/dT^2$ under heating, and cooling respectively are chosen to represent the upper limit of the phase coexistence interval. Similarly, $T_2^h$, and $T_2^c$, corresponding to the minima of the second derivative under heating, and cooling respectively, mark the lower bound of the same interval. The area $A$ of the thermal hysteresis loop has been shaded in the upper frame.  Cooling and heating curves are represented in blue and red, respectively.
} \label{fig:MT}
\end{figure}

\begin{figure}[h!]
\includegraphics[width=\columnwidth]{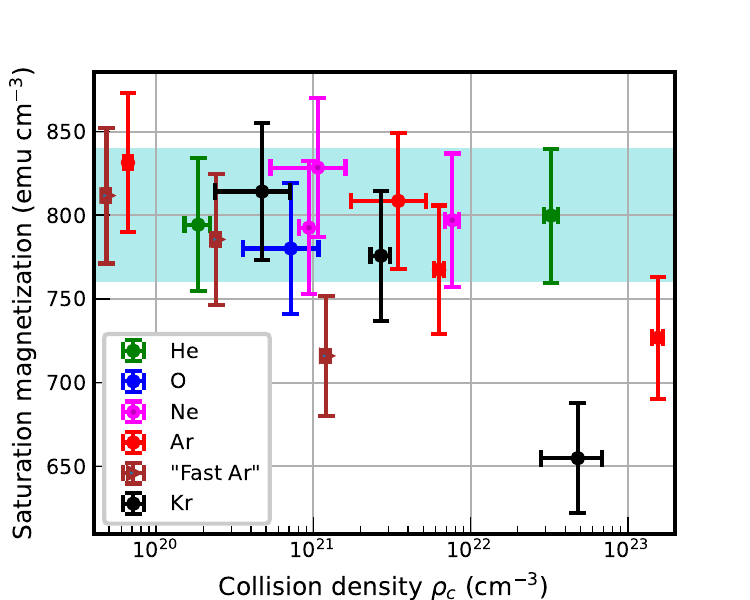} 
\caption{\label{fig:sat} Saturation magnetization (measured with $H=1$~T at 100~K) as function of the collision density}
\end{figure}

The two main modifications observed after ion-irradiation are hysteresis reduction, and the loss of sharpness of the transition, often referred to as \textit{rounding} \cite{Imry1979-1} (see Fig.\ref{fig:MT}). 
To investigate them, isofield thermomagnetic $M(T)$ curves have been measured at constant field $H = 1$~T by a SQUID device 
with $H$ parallel to
the in plane easy magnetization direction. 
This makes the proportionality between $M(T)$ and the ferromagnetic $\alpha$-phase a reliable assumption even under the application of a relatively small field of $1$~T.
Before each measurement, the material has been heated up to $350$~K in order to 
erase any trace of the low-temperature phase, resetting the sample history \cite{Bratko2012-1}. 
Hysteresis, transition temperatures, and rounding are extracted from the $M(T)$ curves and from their derivatives. Whereas hysteresis, and its reduction are phenomena taking place close to the transition temperature $T_t$, the transition rounding is associated with a widening of the phase coexistence region far away from it. 
Here, the thermal hysteresis width, $\Delta T_{hyst}$, is evaluated from the area $A$ between the $M(T)$ heating and cooling curves (Fig.~\ref{fig:MT} top) normalized to the saturation magnetization $M_s$ measured at 100~K with $\Delta T_\mathrm{hyst} = A / M_s$. This is an approach commonly used in the field of loss, and hysteresis modelling; it amounts to consider the effective width of an equivalent rectangular hysteresis cycle \cite{Bertotti1985-1, Bertotti1998-1}.
Indeed, most samples have been irradiated within regimes where the maximum magnetization $M_s$ is not, or is marginally modified, as visible in Fig.~\ref{fig:sat}. 
From the almost unchanged value of $M_s$ and from previous x-ray diffraction characterisations of the film structure after irradiation with neon ions up to a fluence equal to $1.5\times 10^{15}$~ions/cm$^2$ \cite{Trassinelli2014}, no amorphisation of the samples is expected.

Concerning the evaluation of the rounding, the data analysis has to disentangle the effect due to defects-induced  disorder \cite{Vidal2006, Mosca2008} from the strain-induced broadening at the magnetostructural phase transition. 
The latter is driven by the above-mentioned mesoscopic phase-coexistence domain pattern. The former is relevant far from $T_t$ and is  associated with the presence of minority-phase regions whose stability is made possible by a local shift of $T_t$ \cite{Imry1979-1}. 
To tag the irradiation rounding effect only, we introduce the variable 
\begin{equation} \label{eq:delta_T_rounding}
    \Delta T_\mathrm{rnd} = \mathrm{FWHM}_\mathrm{IRR} - \mathrm{FWHM}_\mathrm{REF}
\end{equation}
obtained by the width at half maximum ($\mathrm{FWHM}_\mathrm{IRR}$) of the first derivative of $M(T)$ of the irradiated films minus the contribution $\mathrm{FWHM}_\mathrm{REF}$ of the pristine sample (see Fig.~\ref{fig:MT} bottom). 
The second derivative $d^2M/dT^2$ is used to identify the main part of the phase coexistence interval bounded by
the maximum (at the temperature $T_1$) and the minimum (at $T_2$) of $d^2M/dT^2$ respectively. 
Some typical irradiation-induced modifications are presented, as an example, in Fig.~\ref{fig:MT} where the extrema of $d^2M/dT^2$ under  heating ($T^h_1$, and $T_2^h$), and cooling ($T^c_1$, and $T_2^c$) are marked over the $M(T)$ of the pristine, and of an irradiated sample.

\subsection{Hysteresis: Implanted ions vs induced collisions}

In order to disentangle the role of binary-collisions-induced defects, from the one of ion implantation in modifying the target properties, 
we first compare sample features with respect to the average collision density or to the implanted ion density, $\rho_c$, and $\rho_i$, respectively. 
These quantities are estimated using the applied ion fluences, and the output of the ion-matter interaction Monte Carlo code SRIM/TRIM \cite{Ziegler1985,Ziegler2010}.
More precisely, the outputs from the \emph{Detailed Calculation with Full Damage Cascades} mode are considered, which include the position and number of the produced defects. 
$\rho_c$ is obtained from vacancies and interstitial defects.
The two densities induced by irradiating a sample of thickness $t$ with a fluence $\Phi$ are estimated through the following expressions:  $\rho_i=f \Phi / t$, and  $\rho_c = N_\mathrm{coll}  \Phi / t$, where $f$, is the fraction of ions stopped, and $N_\mathrm{coll}$ is the average number of binary collisions induced by a single ion in the film.
Note that with this simple approach, we do not take into account the depth dependencies of the implanted ion and induced collision densities, but only their average values. 
For irradiations with slow ions, such inhomogeneities can be very important but are not considered in our analysis.
Such an effect is normally attenuated by the possible migration of defects, which is not considered as well.
More specifically for MnAs thin films, characterised by regular volumetric self-organisation of the different phases, irradiation-induced defects at a specific depth are expected to affect the whole sample thickness.
Despite the well-known SRIM/TRIM trend of slightly  overestimating the number of induced defects \cite{Nastasi,Stoller2013,Nordlund2019},  the predicted agglomerations  (i.e. mixed-up vacancies and interstitials clusters)  are rather similar to the ones obtained by advanced many-body molecular dynamics calculations taking into account collective behavior and temperature effects \cite{Nordlund1999,Calder2010,Kim2012, Nordlund2018,Nordlund2019}, and to the ones observed experimentally as well \cite{Lu2016}.
For our specific cases, $f$, and $N_\mathrm{coll}$ have a range of $[0,1]$ and [50,5000], respectively.
Thus, $\rho_i$ and $\rho_c$ differ by orders of magnitude and strongly depend on the projectile ion characteristics.

\begin{figure}[h!]
\includegraphics[width=\columnwidth]{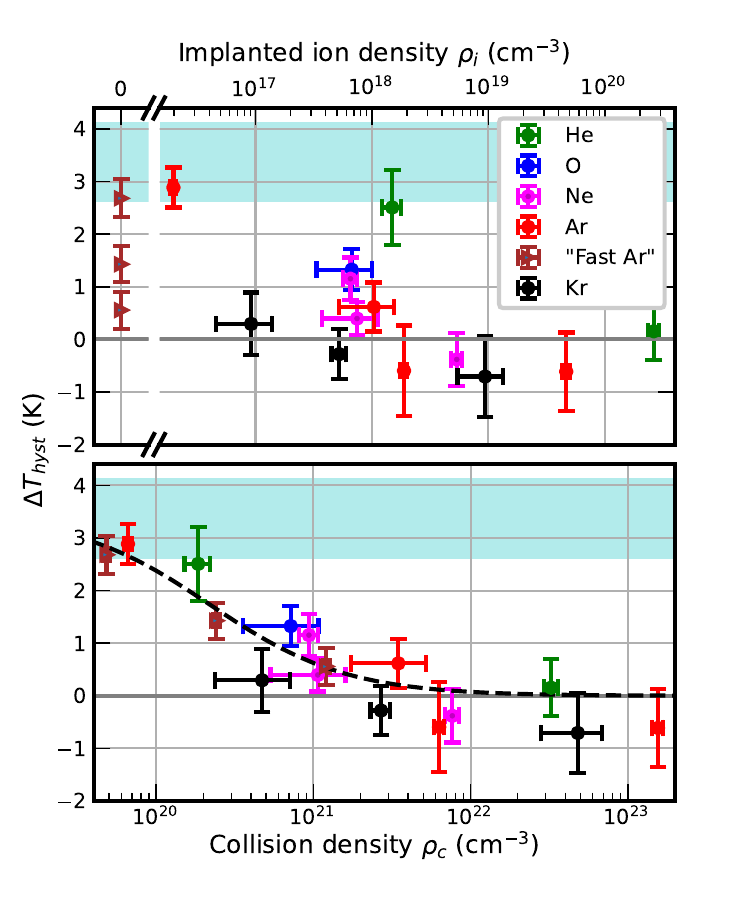}
\caption{Thermal hysteresis area as a function of the average density of implanted ions $\rho_i$ (top), and of the average density of elastic collisions $\rho_c$ (bottom).
In light blue, the value of the pristine sample is reported. The uncertainties on the $x$-axis are mainly determined by the fluence evaluation, which can be critical for some beam preparation. The uncertainty on the $y$-axis is due to the SQUID measurement and the interpolation of the data affected by noise. The dashed line represents the best fit of the data with the formula discussed in the text.
} \label{fig:AvsDion}
\end{figure}

In Fig.\ref{fig:AvsDion} we can see that, whereas no correlation is apparent between $\Delta T_\mathrm{hyst}$ and $\rho_i$, a clear trend can be appreciated when $\Delta T_\mathrm{hyst}$ is plotted against $\rho_c$. More precisely, $\Delta T_\mathrm{hyst}$ decreases as a function of $\rho_c$ until its total suppression at $\rho_c \approx 10^{21}-10^{22}$~cm$^{-3}$. It is noteworthy that the three points relative to fast Ar irradiation, where no ions get implanted into the sample, spread over the same hysteresis decreasing trend curve when plotted against $\rho_c$.
This is one of the main findings of this study, and it shows that the relevant defects are the ones produced by binary elastic collisions. 

\begin{figure}[ht]
\centering
\includegraphics[width=\columnwidth]{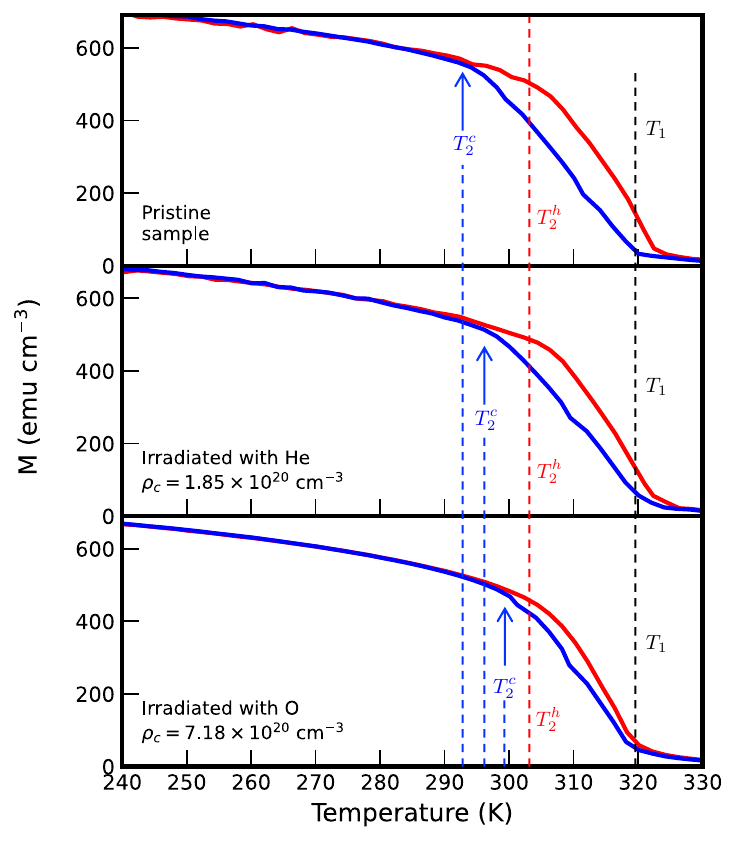}
\caption{
Comparison between $M(T)$ loops over heating, and cooling (in red and blue), measured on the pristine sample (top), on a He irradiated sample (middle) and on an O irradiated sample (bottom). As apparent from Fig.~\ref{fig:delta_T}, $\Delta T_1$ is rather small and does not show any clear trend as a function of the collision density. Thence, a single dashed black line is used to mark $T_1$ over all the curves. On the contrary, $\Delta T_2$ reduces as a function of $\rho_c$. Furthermore, this reduction takes place at fixed $T_2^h$ (i.e. the single dashed red line), with $T_2^c$ getting higher and closer to $T_2^h$ as a function of the collision density (i.e. as shown by the sequence of blue dashed lines in the three frames of the figure). 
\label{fig:T-star} }
\end{figure}

\begin{figure}[ht]
\centering
\includegraphics[width=\columnwidth]{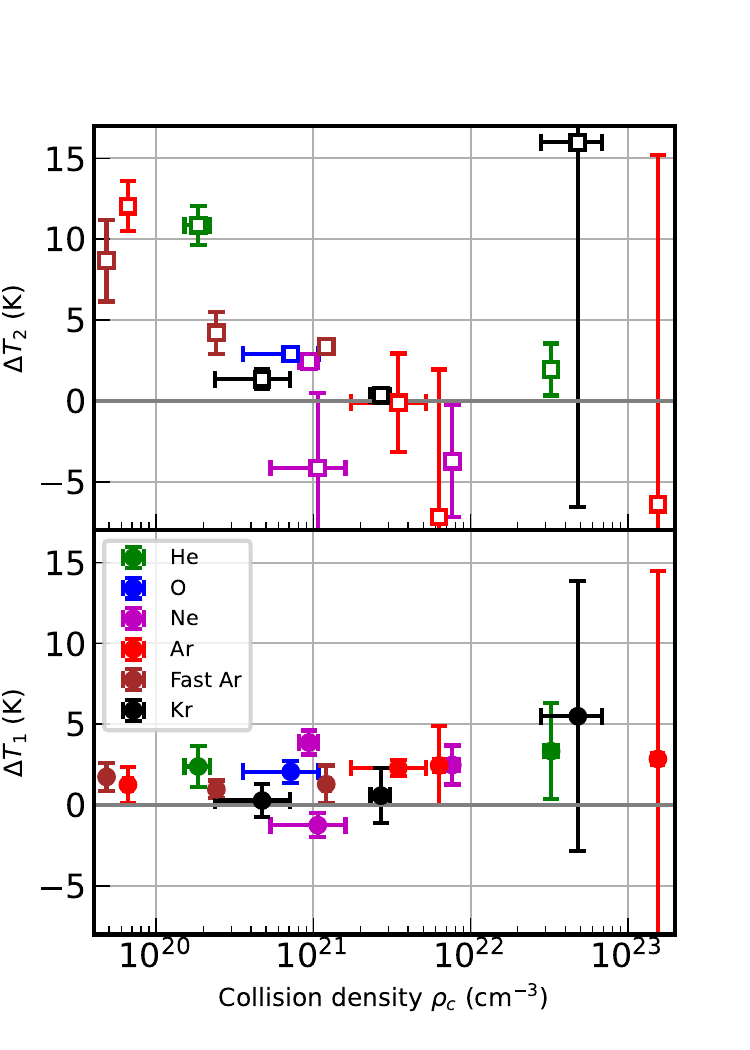}
\caption{ 
Differences $\Delta T_2 = T_2^h - T_2^c$ (top), and  $\Delta T_1 = T_1^h - T_1^c$ (bottom) between the extrema of the coexistence interval  over heating and cooling as function of $\rho_c$. $\Delta T_2 = T_2^h - T_2^c$ reduces as a function of $\rho_c$, similarly to $\Delta T_{hyst}$. $\Delta T_1$ instead is, on average, close to $0$, and does not show any specific trend as a function of $\rho_c$. 
\label{fig:delta_T} }
\end{figure}

\subsection{Hysteresis: additional insights}

To get a better insight into the mechanisms underlying thermal hysteresis in MnAs films, the amplitude of the phase coexistence interval is studied over heating, and cooling as a function of $\rho_c$. When hysteresis is present, the aforementioned second derivative maximum and minimum show different values along the cooling and heating curves (see Fig.~\ref{fig:T-star}, additional curves can be found in \footnote{See Supplemental Material at [URL will be inserted by publisher] for additional magnetisation curves.}).

More precisely, the upper limit of the phase coexistence interval under heating, $T_1^h$, and cooling $T_1^c$ are rather similar, and their difference $\Delta T_1 = T_1^h - T_1^c$, shown in Fig.~\ref{fig:delta_T} (bottom), stays close to zero in all samples, and does not show any specific trend as a function of the collision density $\rho_c$. On the contrary, the difference $\Delta T_2 = T_2^h - T_2^c$ between $T_2$ over heating, and cooling shown in Fig.~\ref{fig:delta_T} (top) is relevant in the pristine sample and gets reduced, and eventually suppressed, as a function of $\rho_c$ following a trend similar to the one of $\Delta T_{hyst}$. In Fig.~\ref{fig:T-star} the values of $T_2^h$, and $T_2^c$ are marked with dashed lines (red and blue, respectively), over the $M(T)$ curves of different samples. In the same figure, a single black dashed line marks $T_1$ over the reference and the irradiated samples. Following \cite{Kaganer2002-1}, $T_1$ can be considered the closest approximation to the actual transition point $T_t$ (i.e. the one of a bulk single crystal). On the other hand, as apparent from Fig.~\ref{fig:T-star}, the reduction of hysteresis takes place as an increase of $T_2^c$, in irradiated samples, with no change of $T_2^h$. In other words, hysteresis disappears through the collapse of the cooling $M(T)$ curve onto the heating one. Hence, hysteresis is mainly associated with supercooled metastable states. 




\begin{figure}[h!]
\includegraphics[width=\columnwidth]{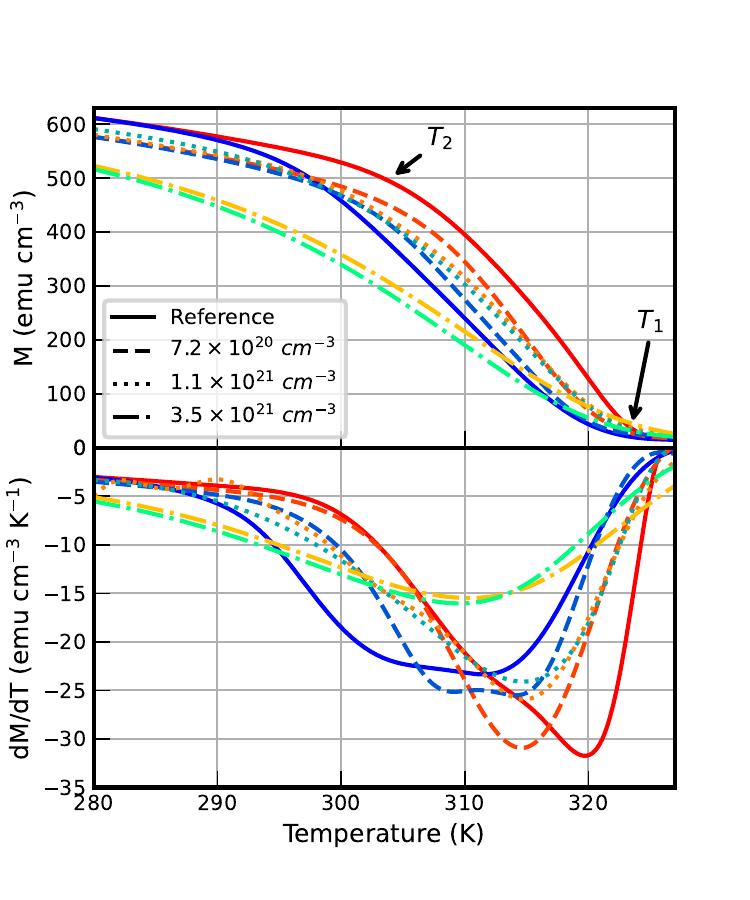}
\caption{
Isofield ($H = 1$~T) magnetization curves $M(T)$  (top) and their first derivative (bottom) measured on reference (continuous lines) and irradiated (dashed lines) samples.
Warm and cold colors correspond to heating and cooling curves, respectively.} \label{fig:MT2}
\end{figure}

Hysteresis is suppressed by ion-irradiation  within the collision density interval  $3 \times 10^{20}$~cm$^{-3} \lesssim \rho_c \lesssim 10^{21}$~cm$^{-3}$ (see Fig.~\ref{fig:AvsDion}).
However, as clearly put in \cite{Kaganer2002-1}, thermal hysteresis takes place in a temperature interval where the material behavior is driven by the self-organized phase-coexistence pattern. A state of affair that can be hardly explained  through a droplet-nucleation mechanism.  

While only the cooling $dM/dT$ curve plotted in Fig.~\ref{fig:MT} (bottom) for the non-irradiated sample shows a good agreement with the linear phase-evolution picture described in Refs.~\cite{Kaganer2000-1, Kaganer2002-1}, all the other curves show a non-linear $M(T)$. Besides, Fig.~\ref{fig:MT} (bottom) and Fig.~\ref{fig:MT2} show that 
the hysteresis reduces jointly with the loss of linearity of the cooling $M(T)$ curve with higher values of the maximum of $dM/dT$ implying sharper phase kinetics as a function of $T$. Similar asymmetries are a rather common feature of first-order phase transitions, from the best-known example of the water-ice transition to  the  different kinetics observed under cooling, and heating in FeRh films \cite{Uhlir2016-1}. Here, as noticeable in Fig.\ref{fig:T-star}, the difference between the cooling, and heating $M(T)$ curves depends on the lower slope of the formers, and on the entailed lower $T_2$ (i.e. a coexistence region lasting till a lower temperature). The loss of linearity and the increasing slope, going hand-in-hand with the hysteresis reduction, take place where the kinetics is led by the stripes domain elastic field. Therefore, all these modifications can be interpreted as a modification of the elastic field associated with the stripes. 


A possible clue to explain this behavior comes from the observation of the distortion of the regular stripes pattern in MnAs films after ion irradiation \cite{Trassinelli2014}, with an increasing density of finite-length stripes. These defects have been observed in non-irradiated samples too \cite{Plake2003-1, Daweritz2003-1, Breitwieser2009-1}. 
They are associated to the non-linear region of $M(T)$ and they mostly disappear within the ordered-stripes structure getting closer to the transition temperature. 
Finite-length stripes may behave as topological defects, modifying the elastic energy with their stress field, similarly to the way unbound edge-dislocations do in 3D solids \cite{Landau1986-1, Nabarro1952-1}, and in 2D mesomorphic phases \cite{Kleman2003-1}. Their increased density in irradiated samples can be the origin of the suppression of the supercooled metastable states, either by changing the phase fraction through nucleation of pairs or by renormalizing the effective elastic constants. 

This explanation, in spite of its being still rather conjectural, raises the question of the way ion-induced disorder directly modifies the phase-domain patterning where the relevant scale is the phase-domain characteristic length (i.e. the stripes spacing, $d_s\approx 0.73~\mu$m \cite{Trassinelli2014}). 

It is worth noting that, as shown in \cite{Kaganer2000-1, Kaganer2002-1}, the stripe phase-domains in MnAs show a regular spacing driven by the substrate elastic field, and fully determined by the film thickness. From this standpoint, the appearance of an increasing number of topological defects within the pattern after ion irradiation, as reported in \cite{Trassinelli2014}, is the signature of a global modification of the pattern involving the full thickness of the film. Putting it another way, hysteresis reduction is associated with a change in the stripes pattern behaving as a 2D mesostructure, notwithstanding the depth distribution of the collision induced impurities.

Over such a large scale, the average collision density $ \rho_c  \propto p$, with $p$ the probability to get a lattice site occupied by some sort of impurity, is expected to be the relevant quantity determining the topological defects density. Nonetheless, some preexisting defect density, associated to an intrinsic impurity density $p_0$ present in the reference sample, must be considered. The dashed curve in Fig.~\ref{fig:AvsDion} (bottom), describing the hysteresis reduction as a function of the collision density $\rho_c$, is obtained using the following fitting function:
\begin{equation}
    f(\rho_c) \propto \frac 1 {p_0 + p} = \frac 1 {p_0 + a \ \rho_c / \rho_0},
\end{equation}
where $\rho_0=2.92 \times 10^{22}$~cm$^{-3}$ is the atomic density of MnAs, $a$ is a proportionality constant between $p$ and $\rho_c/\rho_0$ that incorporates possible auto-healing processes, not taken into account by SRIM/TRIM, and possible activation thresholds, and $p_0$ is a constant describing the impurities present in the sample before the irradiation.
When the value $a = 9.7\%$, determined from the rounding analysis (see next section), is used, the probability value  $p_0=(7.3\pm2.3) \times 10^{-4}$ for the impurity density in the pristine sample is found. This relatively low value is consistent with the typical values of molecular beam epitaxial samples \cite{Chan1995}, and with its nearly defect free stripe pattern reported in \cite{Trassinelli2014}. 
It shows that, besides reducing hysteresis, collision-induced defects can be used to tailor the film domain pattern in the phase coexistence region.

Considering $\rho_c$ as a figure of merit is equivalent to considering the total volume $\langle V_\mathrm{cas} \rangle$ of the ion-induced collisional cascades like in Gibbons model \cite{Gibbons1972,Nastasi}.
The collision density is, in fact, proportional to $N_\mathrm{coll} \phi / t$, where $t$ is the thickness of the sample, $\phi$ is the ion fluence, and $N_\mathrm{coll}$ is the number of collisions per ion, expected to be proportional to the single ion cascade $\langle V_\mathrm{cas} \rangle$.
Due to the relatively small value of $\rho_c$ corresponding to the hysteresis disappearing, effects due to cascade overlaps can be excluded.


\subsection{Rounding}

\begin{figure}[ht]
\includegraphics[width=\columnwidth]{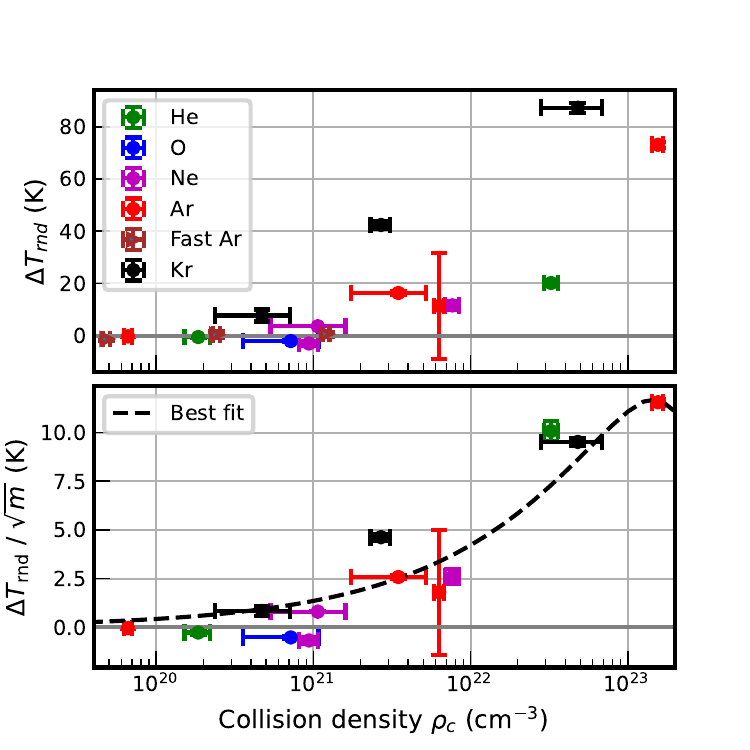}
\caption{Relative rounding $\Delta T_\mathrm{rnd}$ from the FWHM of $dM/dT$ curves without (top) and with $\sqrt{m})$ mass rescaling (bottom). The dashed line represents the best fit of the data with the function $f(p) = c \sqrt{p(1-p)}$ (see text).} \label{fig:rounding_FWHM}
\end{figure}

The transition rounding $\Delta T_\mathrm{rnd}$, deduced from  Eq.\eqref{eq:delta_T_rounding}, also shows a systematic trend as a function of the collision density $\rho_c$ (Fig.~\ref{fig:rounding_FWHM}, top). 
Its  departure from the reference value, at $\rho_c \approx 4 \times 10^{20}$~cm$^{-3}$, is followed by a strong increase with a slope depending on the ion type used for irradiation. Furthermore, the points collapse over the same curve  when normalizing the relative rounding to the square root of the ion mass number $m$, $\Delta T_\mathrm{rnd}/\sqrt{m}$ (Fig.~\ref{fig:rounding_FWHM}, bottom).

Differently from hysteresis suppression, which takes place near $T_t$ where stripes domains and their long-range elastic field drive transition kinetics at the mesoscale \cite{Kaganer2002-1}, rounding takes place far from $T_t$ where the only relevant  scale is the phase transition correlation length $\xi$. Indeed, the rounding behavior fits quite well within the approach reported in \cite{Imry1979-1}.

Considering a magnetic lattice model with a probability $p$ to get a lattice site occupied by some sort of impurity, it is shown that, even when $p \ll 1$, a relevant transition rounding can appear due to the local fluctuations of impurity density. For a random impurity distribution, density fluctuations are proportional to the variance $p (1 - p)$. The key issue to get a stable rounding is the relationship between the phase transition correlation length $\xi$, and the characteristic size $L$ of the regions where  the impurity density fluctuation makes the minority phase 
energetically favored. 
Therefore, the smearing-out of the transition is driven by the dimensionality $d$ of the magnetic system, and by the way the interface energy between the two phases scales with respect to the region size $L$. Following \cite{Imry1979-1}, a spin lattice of dimensionality $d = 3$, with discrete symmetry, and with a surface tension between phases over a region of size $L$ scaling as $L^2$ is expected to be unaffected by quenched disorder when $L < \xi$. Rounding appears when $L \sim \xi$, and scales as $\sqrt{p (1-p)/L^3}$ when $L > \xi$. Assuming the probability $p$, to get an impurity to be $p \propto \rho_c$, from Fig.~\ref{fig:rounding_FWHM} it can be deduced that $\xi \sim L$ when $\rho_c \approx 4 \times 10^{20}$~cm$^{-3}$. 
From cascade simulations, one can observe that one single ion forms several clusters of defects.
The typical volume of collision clusters is $\propto 1/m$, where $m$ is the ion mass. Assuming elastic-collision cascades as the origin of the impurities implies $L^3 \propto 1/m$, a feature that explains the higher slopes shown by samples irradiated with heavier ions and the collapse of all the points onto the same curve after the $1/\sqrt{m}$ normalization.
Similar curves can be obtained by normalizing with respect to the number of collisions per ion and the average size of clusters of defects.
Figure~\ref{fig:rounding_FWHM} (bottom) shows the mass normalized relative rounding data fitted with a function 
\begin{equation}
f(\rho_c) = c \sqrt{a \frac{\rho_c}{\rho_0} \left(1- a \frac{\rho_c}{\rho_0}\right)} \equiv c \sqrt{p (1-p)}.
\end{equation}
From this analysis, a value of $a = (9.7 \pm 0.6) \%$ is found.

This shows that far away from the transition temperature, the minority phase 
develops as randomly distributed droplets before the building-up of the elastic-domain stripe structure as reported in \cite{Plake2002-1, Daweritz2003-1}. 
Whereas closer to $T_t$ energy minimization is dominated by the long-range elastic interaction between stripes \cite{Kaganer2000-1}, in the first stages of the transition the leading energy terms are local, and fully led by the size $L$ of regions where the precursor develops. 

For the high-fluence regime considered for the rounding effect, one could also consider the effect of the collisional cascade overlapping.
When we consider the simple model from Gibbons \cite{Gibbons1972,Nastasi}, the fraction of the irradiated volume $V_I$ over the total sample volume $V_0$ is given by
\begin{equation}
\frac {V_I}{V_0} = 1 - e^{-\frac{\langle V_\mathrm{cas} \rangle \phi}{ t}} = 1 - e^{- C \rho_c},
\end{equation}
with $\langle V_\mathrm{cas} \rangle$ the average cascade volume, $\phi$ the ion fluence, and $t$ the sample thickness.
Introducing an appropriated constant $C$, this formula can be written in terms of collision density $\rho_c$ (see previous section) and thus should be independent of the ion type.
However, this approach does not consider the cascade fragmentation into defect clusters that is taken into account in our analysis by the $\sqrt{m}$ dependence.
Except for the very high density values of $\rho_c$, most of the data described here are expected to be in the regime with $p =a \rho_c/\rho_0 \ll 1$, where cascade overlapping can be neglected.

\section{Conclusions}

In this work, we address the changes induced in MnAs thin films by ion irradiation. It is shown that elastic ion-atom collisions are the main mechanism responsible for modifying the magnetic properties of the material. Besides, modifications induced by ion collisions are shown to act at two different scales. In the short range, collisions serve as nucleation seeds of precursors responsible for the loss of sharpness of the phase change far from the transition temperature. At a larger scale, collisions modify the elastic energy by softening the long-range elastic field associated with the substrate coupling.
Thus, the creation of new topological defects is favored in a distorted stripe pattern, and the thermal hysteresis is reduced until its elimination. 

Disorder induced by light ions irradiation of magnetic thin films has been recently investigated through domain wall motion measurements focusing on spintronics applications \cite{vanderJagt2022-1}. Here, the effect of irradiation with ions of different mass is investigated through the behavior of the first-order magneto-structural transition on MnAs thin epitaxial films. Hysteresis suppression, transition rounding, and the tailoring of the phase-domain pattern are observed, opening unique opportunities in terms of applications to magnetocaloric devices, and to design of novel magnetic heterostructures. 

\vspace{5mm}
\hfill

\begin{acknowledgments}
The authors acknowledge the staff (and in particular D. Hrabovsky) of the MPBT (physical properties - low temperature)
platform of Sorbonne Université for their support.
This work was supported by French state funds managed by the ANR within
the Investissements d’Avenir programme under reference ANR-11-IDEX-0004-02, and
within the framework of the Cluster of Excellence MATISSE led by Sorbonne Universit\'es, and of the project HiPerTher-Mag (ANR-18-CE05-0019).
\end{acknowledgments}

\bibliography{MnAs_irr}

\vspace{10cm}


\pagebreak
\newpage

\begin{center}
\textbf{\large Supplemental Materials}
\end{center}
\setcounter{equation}{0}
\setcounter{figure}{0}
\setcounter{table}{0}
\makeatletter
\renewcommand{\theequation}{S\arabic{equation}}
\renewcommand{\thefigure}{S\arabic{figure}}
\renewcommand{\bibnumfmt}[1]{[S#1]}
\renewcommand{\citenumfont}[1]{S#1}

\section{Details of the irradiation conditions anad characterization measurements}

\onecolumngrid\
\begin{figure}
\includegraphics[trim={20 80 40 120},clip,width=0.7\columnwidth]{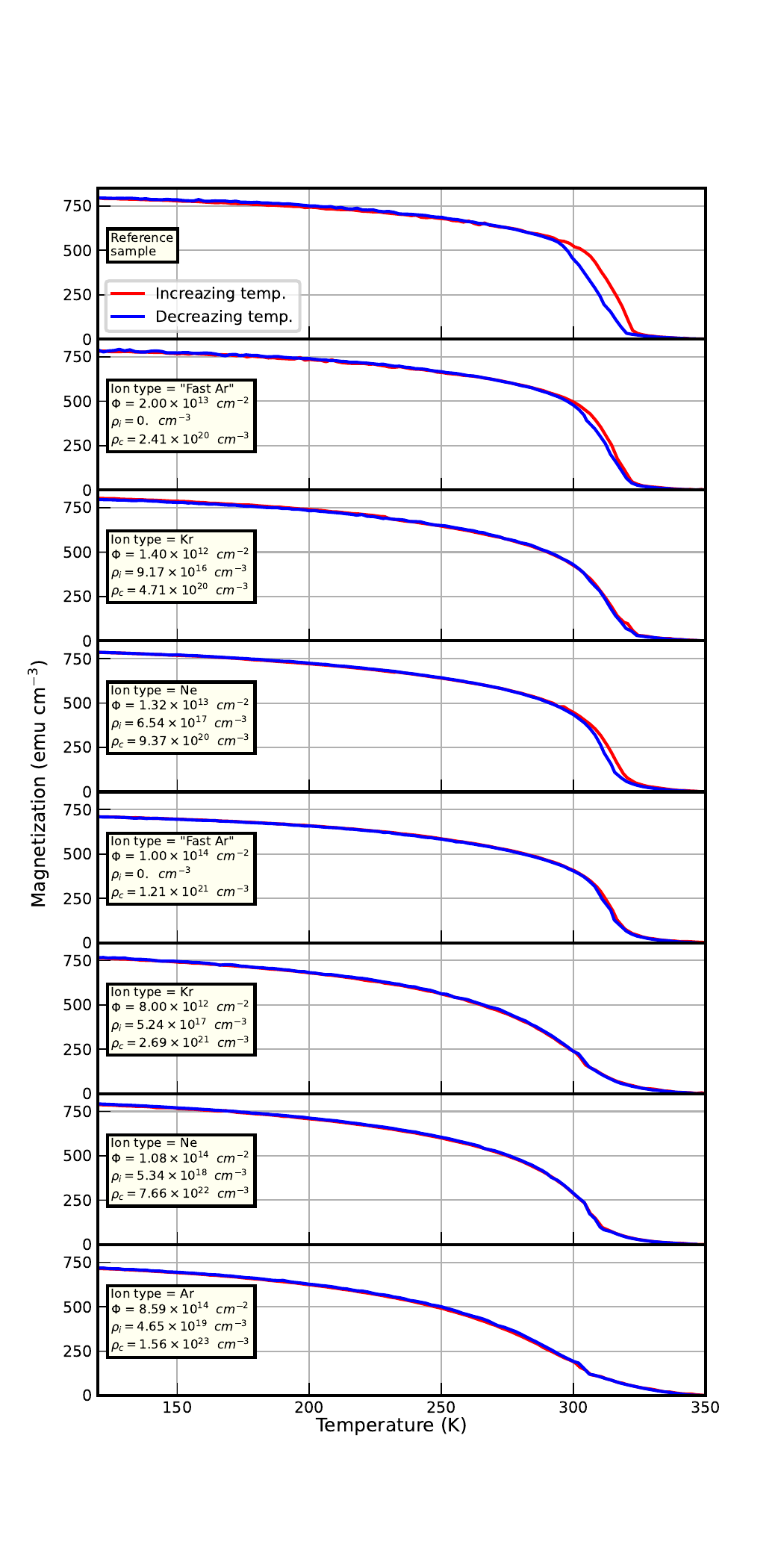}
\caption{
Isofield ($H = 1$~T) magnetization curves $M(T)$ of selected samples. Irradiation details are indicated in the corresponding subplots.
\emph{Fast ions} indicate irradiations with high kinetic energy ions. Irradiation details can be found in Table~\ref{tab:data}.} \label{fig:MT_all}
\end{figure}
\twocolumngrid

\end{document}